# Advances in Shell Buckling: Theory and Experiments

## J. Michael T. Thompson

***Department of Applied Maths & Theoretical Physics,
University of Cambridge, CB3 0WA, UK***

## Abstract

In a recent feature article in this journal, co-authored by Gert van der Heijden, I described the static-dynamic analogy and its role in understanding the localized post-buckling of shell-like structures, looking exclusively at integrable systems. We showed the true significance of the Maxwell energy criterion load in predicting the sudden onset of 'shock sensitivity' to lateral disturbances. The present paper extends the survey to cover non-integrable systems, such as thin compressed shells. These exhibit spatial chaos, generating a multiplicity of localized paths (and escape routes) with complex snaking and laddering phenomena. The final theoretical contribution shows how these concepts relate to the response and energy barriers of an axially compressed cylindrical shell.

   After surveying NASA's current shell-testing programme, a new non-destructive technique is proposed to estimate the 'shock sensitivity' of a laboratory specimen that is in a compressed meta-stable state before buckling. A probe is used to measure the nonlinear load-deflection characteristic under a rigidly applied lateral displacement. Sensing the passive resisting force, it can be plotted in real time against the displacement, displaying an equilibrium path along which the force rises to a maximum and then decreases to zero: having reached the free state of the shell that forms a mountain-pass in the potential energy. The area under this graph gives the energy barrier against lateral shocks. The test is repeated at different levels of the overall compression. If a symmetry-breaking bifurcation is encountered on the path, computer simulations show how this can be supressed by a controlled secondary probe tuned to deliver zero force on the shell.

## 1. Introduction

   Early in the 20[th] century the pioneering use of thin metal shells as load-carrying components in aircraft and rockets stimulated engineers to look in detail at two well-defined archetypal problems of elastic buckling. These were the complete spherical shell subjected to uniform external pressure, and the cylindrical shell subjected to uniform axial compression. In careful laboratory tests, both of these were found to be collapsing violently at about one quarter of the classical buckling loads, $P_C$, predicted by small-deflection linear theory. In response to this discrepancy, Karman & Tsien [1939, 1941] made approximate

Rayleigh-Ritz analyses to demonstrate that, for both problems, there exists a very unstable, sub-critical post-buckling path of periodic equilibrium states. This falls rapidly from $P_C$ and eventually stabilises at a fold (limit point) at what they termed the *lower buckling* load, $P_L$. They suggested that this load might be a useful empirical 'lower bound' for the collapse load of real shells which would certainly have inevitable imperfections and finite disturbances.

Subsequently Tsien [1942] focused attention on a result of Friedrichs [1941] who showed that just above $P_L$ there was a load, $P_M$, at which the grossly deformed but stabilized path first had a total potential energy less than that of the trivial unbuckled state. Akin to Maxwell's criterion that a thermodynamic state is likely to be found in the state of minimum energy, Tsien seized on $P_M$ as the *energy criterion* load. Uncritically, and without any real evidence, he decided that this would be a logical failure load due to dynamic disturbances. Given that the experiments were performed in careful laboratory tests where disturbances were minimal, this explanation could not be true, since it would imply that near $P_M$ the shell would be constantly jumping in and out of its buckled state. Tsien later retracted his view [Tsien, 1947], though many researchers probably never noticed this paper, and the relevance of $P_M$ was still being discussed (and convincingly disproved) twenty years later by Babcock [1967].

It is a new understanding of the energy criterion load (now called the Maxwell load), as signifying the onset of 'shock sensitivity' that I present in this paper following Thompson & van der Heijden [2014]. Also presented is a proposal for a novel non-destructive experimental approach to assess this sensitivity, sketched in Fig. 1.

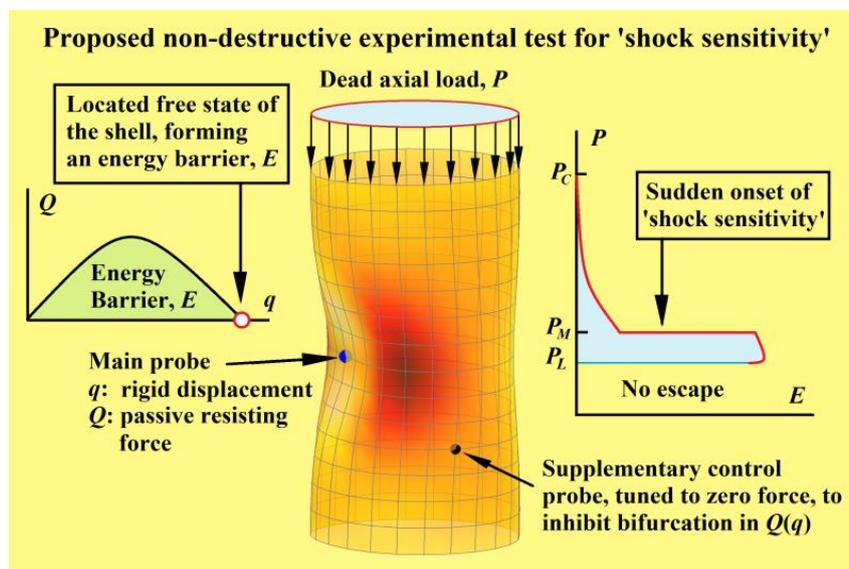

Fig. 1 A sketch of the proposed experimental procedure, described more fully in §9. The displayed shell-like structure is the result of a preliminary computational study by Jan Sieber, and shows the main probe on the left, and a supplementary probe on the right which was needed to suppress a pitch-fork bifurcation in the lateral load-deflection response, $Q(q)$.

This 'shock sensitivity' is important in its own right, but it is not seen as explaining the low experimental failure loads. These are largely due to the imperfection-sensitivity of the shells to initial geometrical imperfections in the shape of the middle-surface, as described in the ground-breaking thesis of Koiter [1945]. Meanwhile, some different approaches and points of view are given by Croll & Batista [1981], Yamada & Croll [1999], Zhu *et al* [2002] and Elishakoff [2012].

A full literature review relevant to the material presented here, and running from von Karman & Tsien [1939, 1941] to the present day is given in [Thompson & van der Heijden, 2014]. Meanwhile an excellent overview of the wider shell buckling field is given in the web-page of Bushnell [2014].

## 1.1 New theoretical concepts

In an earlier feature article published in the *International Journal of Bifurcation and Chaos*, we gave an introduction to the static-dynamic analogy and its role in understanding the post-buckling responses of shell-like structures [Thompson & van der Heijden, 2014]: this looked exclusively at the behaviour of integrable systems. In particular, it showed a true consequence of the energy criterion load (now called the Maxwell load) in predicting the onset of 'shock sensitivity'. The present paper can be seen as a continuation of this earlier one, extending it to cover non-integrable systems (such as the compressed cylindrical shell), which exhibit spatial chaos, manifesting itself as snaking and laddering of the post-buckling path. To make the current paper reasonably self-contained, I give in §2 and §3 a brief summary of this earlier contribution before tackling the new material. In §4 we describe the spatial chaos and multiplicity of localized paths that accompany non-integrability, before giving in §5 an outline of snaking re-stabilization in non-integrable systems. In §6 we look at bifurcations on the snaking paths which give rise to short (asymmetric) linking paths, akin to the rungs of ladders. The final theoretical survey is to see how these new concepts relate to the post-buckling response of a long axially compressed cylindrical shell.

All of these theoretical advances draw on the wonderful progress that has been achieved in recent years by the Bath and Bristol groups, [Hunt *et al*, 1989; Hunt & Lucena Neto, 1993; Champneys *et al* 1999; Hunt *et al* 2000, 2003; Budd *et al* 2001; Horak *et al* 2006].

## 1.2 NASA tests and controlled experiments

The second half of this paper is devoted to laboratory testing of shells, starting with a brief review of the historical data on the premature scatter of experimental buckling loads for the axially compressed cylinder in §8.1. This is followed in §8.2 by a quick look at the current NASA programme of full scale tests on unwanted shells left over from the space shuttle era.

The main experimental feature, in §9, is then an outline of a proposed new testing technique to estimate the shock sensitivity uncovered in the latest theoretical work. The aim is to develop a non-destructive experimental testing procedure to determine the shock-sensitivity of a thin elastic shell to (static or dynamic) lateral side-loads. Shells of any shape can be tested, and are presumed to be significantly loaded in a fixed membrane compression so that they are in a meta-stable state (stable for small disturbances, unstable for large).

Using a probe, we aim to measure the nonlinear load-deflection characteristic of the shell under a rigidly applied lateral displacement. Sensing the passive resisting force of the shell, we can in real time plot the encountered load-deflection diagram. This will in general show an equilibrium path that rises to a maximum of the force and then decreases to a state in which the force has dropped to zero. This means that we have located an unstable equilibrium state of the free shell that forms a mountain-pass in the total potential energy functional of the shell. The area under the load-deflection curve gives us the energy barrier that must be overcome by any static or dynamic lateral disturbances that impinge on the shell at the point of the probe. The test can finally be repeated at different levels of membrane compression.

Our proposed experimental work (still in the planning stage) will look first at the axially compressed cylindrical shell for which a lot of background data and concepts are available, especially if the shell is unstiffened and long. But the experimental technique is not restricted in any such way. Possible complications such as a bifurcation in the lateral response are examined in some detail, and computations by Jan Sieber show how these can be overcome by the addition of an extra control probe to stabilize the shell under test. The analogy with a deep elastic arch is explored, and it is shown how a mountain pass in the potential energy of a shell with the shape of a small dimple could allow a dynamic jump to bypass the large energy barrier associated with the unstable overall post-buckling pattern. No such experiment has yet been made, but some equipment is currently being assembled by Lawrence Virgin at Duke University.

Experimentalists who wish to learn about this suggested approach might care to jump ahead to §7 or even §8 since a detailed understanding of the theory is not essential, certainly on a first reading.

## 2. The Static-Dynamic Analogy

### 2.1 Localization as a homoclinic orbit

The twisted isotropic rod (namely a rod with equal bending stiffness in each direction) gives the simplest localization scenario, Thompson & Champneys [1996], van der Heijden & Thompson [2000], and it has a *real* analogy with the symmetric top. Being an integrable system, it gives a usefully simplified introduction to the underlying ideas. Meanwhile, the strut on a nonlinear, quadratic foundation has a *virtual* dynamic analogy: it is non-integrable, but a double-scale perturbation renders it integrable close to the buckling bifurcation. We use these two problems to show how the localized solutions offer an order-of-magnitude lower energy barrier than is offered by the periodic states.

### 2.2 Twisted rod and spinning top

The analogy between twisted rods and spinning tops, first pointed out by Kirchhoff, holds for symmetric and non-symmetric systems, but it is the integrable symmetric cases that concern us in this section.

The behaviour of a circular twisted rod, made of an elastic material, compares precisely with the spinning of a symmetric top. The long elastic rod (deemed theoretically to be infinite in length) is loaded at its ends by a tension, $T$, and a twisting moment, $M$. Its behaviour is governed by the composite *moment* parameter, $m := M/\sqrt{(BT)}$, where $B$ is the bending stiffness about any axis. Meanwhile the symmetric spinning top has a corresponding *momentum* parameter, $m := \alpha /\sqrt{(INgl)}$, which is defined in terms of the angular momentum about the fixed vertical axis, $\alpha$, the moment of inertia of the top about it spin axis, $I$, the mass, $N$, the gravitational constant, $g$, and the standing height of its centre of gravity $l$.

On identifying the independent axial coordinate of the rod, $x$, with the time, $t$, of the top, it is well known that the equations governing the static spatial deformation of the rod are identical to those governing the dynamic motions of the top. For arbitrarily large displacements, both sets of equations can be solved exactly in terms of an equivalent one-degree-of-freedom mechanical oscillator. This oscillator varies with the parameter $m$, and Fig. 2 shows in white two typical phase portraits on a plot of the Euler angle, $\theta$ against its (space or time) derivative $\theta'$.

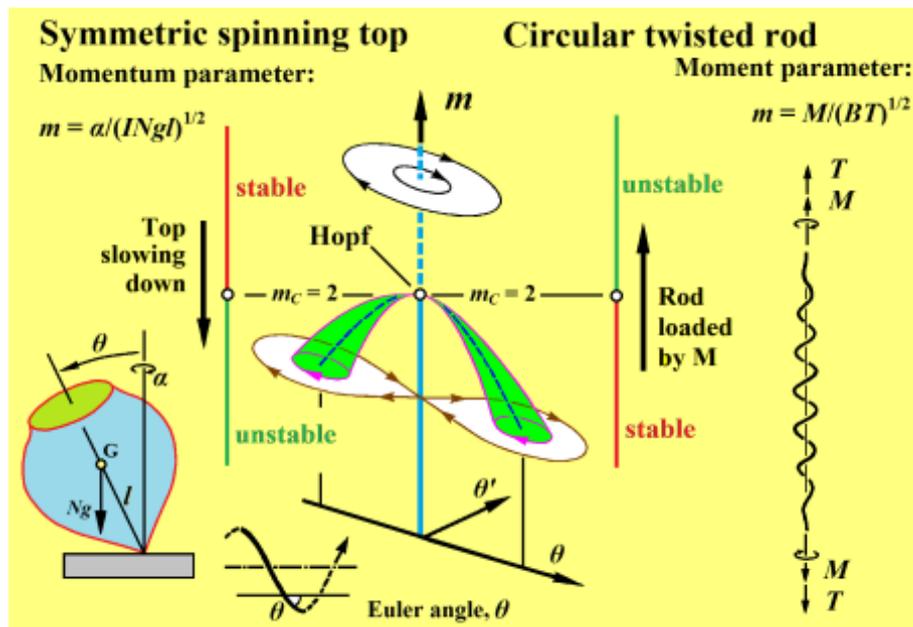

Fig. 2 Illustration of the static-dynamic analogy for a symmetric spinning top and a twisted rod of circular cross-section, showing phase portraits of the equivalent oscillator at two values of the control parameter, $m$.

The portraits change their form at the critical value of $m = m_C = 2$, at which there is (in the terminology of nonlinear dynamics) a super-critical Hamiltonian Hopf bifurcation. Note that it is indeed supercritical for the top, but for the rod appears as a subcritical event.

Under increasing load parameter, $m$, the initially straight rod loses its stability at $m_C$, where linear theory would predict a uniform helical (periodic) deformation. A number of unstable, sub-critical equilibrium paths are generated at $m_C$ as we shall illustrate more fully in Fig. 3. Conversely, the top is stable for $m > m_C$, and under a slow decrease of its rate of spin (implying a decrease of $\alpha$ and therefore $m$) its vertical spinning state becomes unstable at $m_C$.

## 2.3 Equivalent oscillator for a circular twisted rod

Fig. 3 is drawn specifically for the circular twisted rod (meaning a rod of circular cross-section, or more generally any rod with equal bending stiffness in every direction, which includes a square cross-section).

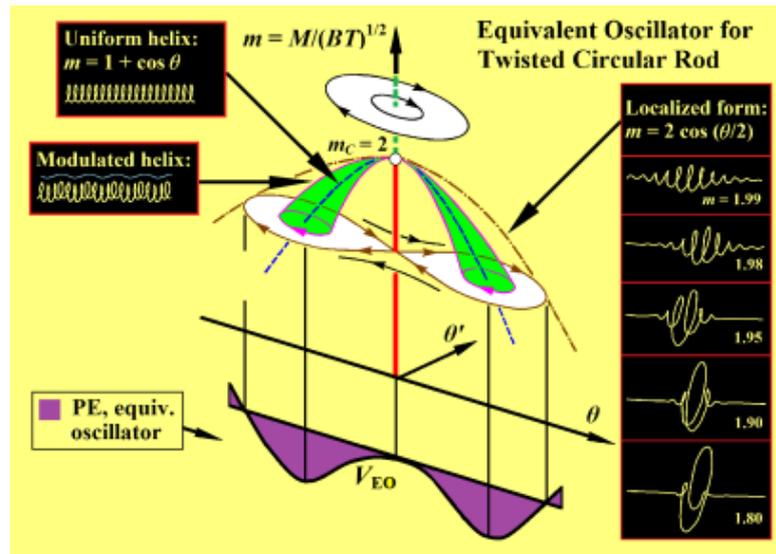

Fig. 3 Response of the stretched and twisted circular rod, showing the total potential energy of the equivalent oscillator, and examples of the helical and localized forms.

Details of the deflected equilibrium states of the rod are shown, relating in particular to the lower white phase portrait of the equivalent oscillator at a load parameter, $m < m_C$. This lower portrait corresponds to a ball rolling on the potential energy of the equivalent oscillator shown in purple. Note carefully that this energy is an artefact of the analysis, and is NOT the potential energy of the physical system. It varies with $m$, such that at $m > m_C$ it has just a central minimum. The equivalent oscillator thus displays the response of a pitch-fork bifurcation. These phase portraits effectively hide the intrinsic helical response, and it is useful to think of $\theta$ as representing a modulation of the intrinsic helix. Thus the curve in blue, which corresponds to fixed points in the phase portraits, represents just a straight-forward helical response of the rod (with fixed $\theta$ and no modulation). The *homoclinic* orbits of the portraits can be thought of as leaving the red trivial solution in infinite time, making a fast loop and then returning infinitely slowly to the *same* solution (hence the use of the adjective homoclinic). Descriptions of homoclinic orbits, and the heteroclinic orbits that we shall encounter later, can be found in [Thompson & Stewart 1986]. This homoclinic orbit corresponds to an un-damped ball rolling on the equivalent potential after being given a minute nudge from the central hill-top. Adding in the intrinsic helical behaviour, these then correspond to the localized solutions shown on the right-hand side. Finally the closed phase orbits around the non-trivial fixed points correspond to equilibrium states in the form of a modulated helix, as drawn.

## 2.4 Energy barriers against lateral disturbances

Many long structures with unstable shell-like post-buckling characteristics exhibit a large number of falling post-buckling paths, like those we have just described for the twisted rod. We will concentrate here on dead loading, in which (for example) $M$ and $T$ are prescribed, rather than rigid loading in which their corresponding displacements are prescribed. All these falling paths are then certainly unstable, and we must therefore enquire about their physical relevance.

To answer this, we consider a general shell-like structure under load $P$. Note in passing that when we are speaking more generally than about a twisted rod, we write the load parameter as $P$ (rather than $m$) and use the wider adjective 'periodic' rather than 'helical'. Below the critical buckling load, $P < P_C$, the trivial state is (meta-) stable and the falling equilibrium paths define mountain passes in the total potential energy. Under static or dynamic lateral disturbances, a structure in its trivial state would have to surmount one or other of these passes if it were to buckle and fail. So the height of these passes, relative to the local minimum in which the unbuckled system rests, is of key importance in ensuring the integrity of the structure.

This is illustrated schematically in Fig. 4, where the total potential energy, $V$, is sketched as a function of the generalized coordinates (mode amplitudes, say), $q_i$, at a value of $P_1 < P_C$.

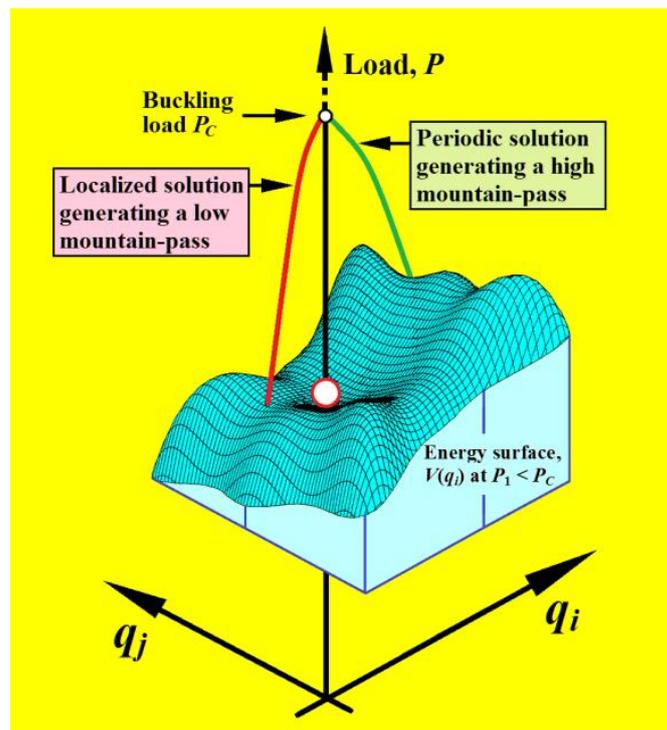

Fig. 4 Schematic of the total potential energy at a fixed value of the load less than the critical, showing the energy passes, of differing heights, corresponding to periodic and localized unstable paths.

We can imagine a small ball resting on this surface at the trivial un-deflected state, and subjected to small disturbances. Escape of the ball, corresponding to buckling of the structure, will depend on surmounting various mountain passes, generated by falling equilibrium paths. Several passes are shown, with two emphasized by the addition of their equilibrium paths. Following what we shall next describe for the twisted rod, we have labelled one as a periodic path and one as a localized path.

## 2.5 Barriers for two integrable systems

Having seen the significance of the energy barriers generated by the unstable falling equilibrium paths, we now look at the calculated values for the twisted circular rod, displayed in Fig. 5.

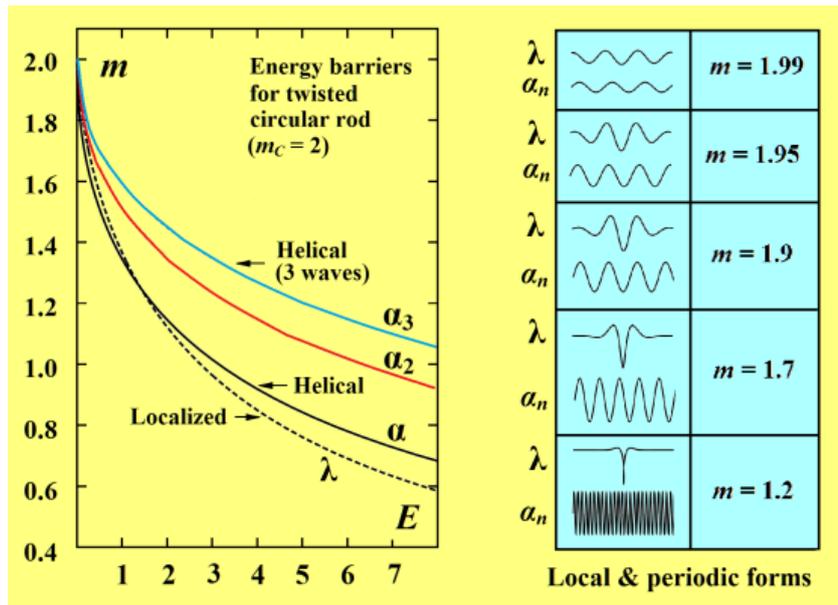

Fig. 5 Calculated energy barriers for the circular twisted rod, comparing the localized and helical barriers. The localized barrier is seen to be roughly equal to the barrier of a single helical wave, which is however a kinematically inadmissible deformation.

The graph on the left shows the variation of the energy barrier, $E$, at different values of the load parameter $m$. For each value of $m$, the barrier is measured from the energy datum of the corresponding (same $m$) trivial un-deflected state.

The aim is to compare the values of $E$ offered by the localized and helical solutions, which are illustrated on the right-hand diagrams. Plotting $m$ versus $E$ for the localized solution is straightforward, and gives the curve $\lambda$. However, when we turn to the helical solutions we must address the fact that the helices extend along the whole length of the supposedly infinite rod. The $E$ for this continuous helix is therefore technically infinite. To overcome this, we have plotted curve $\alpha$ using the energy of a single helical wave, curve $\alpha_2$ for two helical waves (multiplying the $E$ value by two), and $\alpha_3$ for three waves (multiplying the $E$ value by three). As we might expect, curve $\lambda$ is quite close to curve $\alpha_1$, but we must remember that these finite length helical waves do not represent kinematically admissible displacements of the rod.

Now anything treated as 'long' must clearly contain at the very least 5 complete periodic waves (we use the adjective periodic when generalizing the discussion beyond, but still including the helix). So we can conclude that the localized solution offers a barrier that is at least five times lower than that offered by the periodic solution: and will indeed usually offer something even lower. To describe this dramatic reduction in the energy barrier triggered by the localized states we recently coined the expression 'shock sensitivity'.

A second parallel demonstration of this phenomenon is given by the strut on a nonlinear (quadratic) elastic foundation, within a perturbation analysis as discussed in some detail in [Thompson & van der Heijden, 2014]. In this, the first-order perturbation equation gives the first-order solution as $u = A(X) \cos x$ where $u$ is the displacement at distance $x$ along the strut. Here $A$, a function of the 'slow' independent variable, $X$, can be thought of as a slow modulation of the buckling displacement $\cos x$. Then, the third-order equation is that of a one-degree-of-freedom nonlinear oscillator in $A(X)$, remembering that in the static dynamic analogy we constantly jump between viewing $x$ (and hence $X$) as a space or time variable.

The results, given in [Thompson & van der Heijden, 2014], fully confirm the form of curves derived from the twisted rod, displayed in Fig. 5.

## 3. Re-stabilization in Integrable Systems

### 3.1 Lower buckling load of Karman and Tsien

Our major objective in this paper is to examine systems that have the re-stabilization characterized by the buckling of elastic shells, and the post-buckling response of an axially loaded cylindrical shell was shown by von Karman & Tsien [1939, 1941] to have the form of Fig. 6.

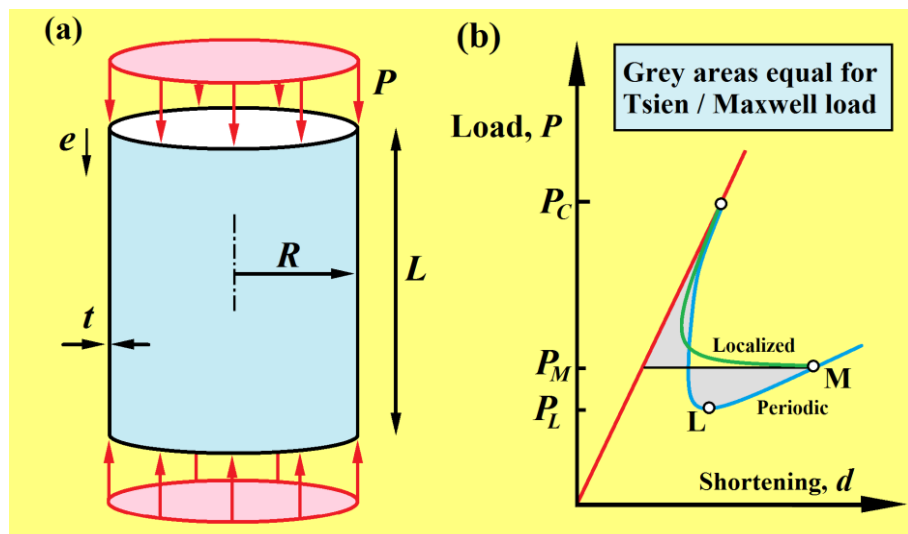

Fig. 6 The classical post-buckling scenario for an axially compressed cylindrical shell, showing the lower buckling load, $P_L$, and the Tsien/Maxwell load, $P_M$. The periodic state at the Maxwell load has the same energy as the trivial solution, corresponding to the two equal grey areas. The energy of the localized solution at the Maxwell load is discussed more fully in Fig. 8. Note that in the above heuristic sketch, the shell post-buckling is drawn as if it were for an integrable system: see Fig. 16 for the real thing.

Here, on the right-hand side, we show the load versus end shortening response of the shell. The red line is the uniform contraction before buckling, and the blue curve represents the periodic solutions obtained by Karman and Tsien. They suggested, as a useful approximation or bound to the premature experimental buckling values (sometimes as low as $P_C / 4$), the *lower buckling load $P_L$.* Then, in a follow-up paper, Tsien [1942] stressed the importance (and even the logic) of what he called the *energy criterion buckling load* at $P_M$ where the total potential energy in the trivial state is equal to the total potential energy in the re-stabilized periodic solution. Tsien [1947] later admitted his error about the logic of $P_M$, but many researchers (perhaps not having seen the admission) continued to use it for quite a number of years. It was solidly repudiated by Babcock [1967], who showed that, contrary to the predictions of Tsien's criterion, there was no observable difference in experimental buckling loads between tests under dead and rigid loading conditions. Unknown at the time, was the localized post-buckling curve sketched, as in [Hunt & Lucena Neto, 1993], in green. Note that this green curve is displayed here as if the shell were an integrable system, as it would be within an approximate energy analysis. We show later that the true significance of $P_M$ (now called the Maxwell load) is that it represents the end of the falling localized path.

There is an apparent anomaly in Fig 6(b), because by areas we can prove that the energy at M is zero (by following the periodic circuit) or non-zero (by the localized circuit). The answer is (at least for an integrable system) that point M really represents two distinct states, the continuous periodic state and the discontinuous localized state which has a transition point from being trivial (straight) to being periodic. This subtle point is clarified later in Fig. 8 in §3.3.

The definition of the Maxwell load that we use throughout the present paper is based on equating the total potential energy (strain energy, plus load energy) of the loaded trivial to that of the loaded stable periodic state. This ties in with some remarks by Mark Peletier (in a personal communication). For the infinitely long cylindrical shell he defines the Maxwell load in two ways which give the same answer. It is the minimum load with negative total potential energy relative to the trivial: over all equilibria; or alternatively over all periodic equilibria. Additionally, depending on our particular interest, we can apply these definitions to a cylindrical shell either globally, over all circumferential wave numbers, or specifically within a prescribed wave number.

### 3.2 Re-stabilization of rod in a tube

So heading towards our goal of better understanding shell post-buckling, we look now at re-stabilization, but first for an integrable system. The twisted isotropic rod constrained to deform on a cylinder is a good example [van der Heijden 2001]. For simplicity, we will call this a rod *in* a cylinder or tube (choosing the latter to avoid confusion with the shell), but depending on the direction of the constraining pressure it might apply alternatively to a rod on the outside of the cylinder or tube.

We shall see how the falling localized post-buckling solution is destroyed at a heteroclinic connection between two *different* saddles [Thompson & Stewart 1986] at the Maxwell load. This means that *shock-sensitivity* starts at the Maxwell load, giving now a correct logical foundation to Tsien's energy criterion load.

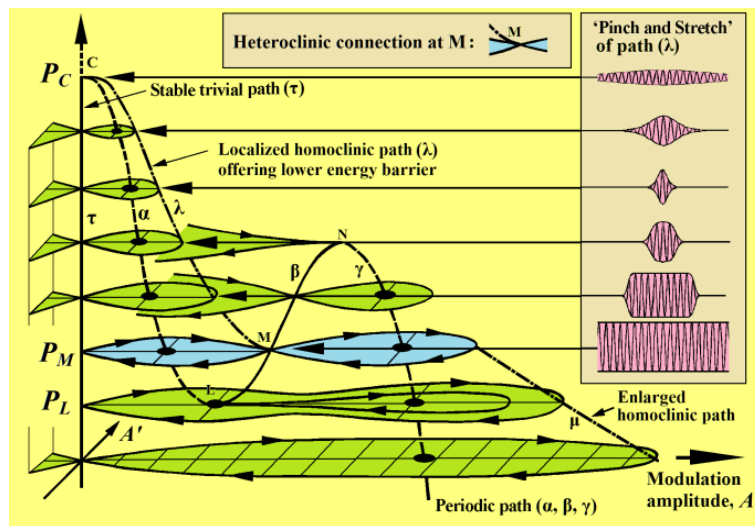

Fig.7 Schematic of the response of an integrable system with restabilization of the periodic post-buckling path. A heteroclinic saddle connection in the response of the equivalent oscillator at $P_M$ destroys the falling localized path.

The diagram of Fig. 7 illustrates the behaviour of the elastic circular rod in a cylinder or tube [van der Heijden, *et al, 2002*]. The most dramatic liberty that we have taken in drawing this schematic picture is that for the cylindrically constrained rod the critical load $P_C$ is actually infinite. We have sketched it at finite $P$ for convenience in comparing the rest of the picture with what we want to understand about re-stabilization. Once again this integrable system has an equivalent nonlinear oscillator with one degree of freedom, and its phase portraits are sketched for different values of the load $P$. We now have a periodic path which falls to $P_L$ where it re-stabilizes at L before falling again after the local maximum, N. A localized path is apparent from the homoclinic orbits of the oscillator, and we see that this localized path vanishes on collision with the re-stabilised periodic path at M, at the Maxwell load, $P_M$. This corresponds to a heteroclinic connection of the oscillator, at which a slightly perturbed solution close to the trivial state moves in 'infinite' time to a point close to the unstable periodic solution.

The diagrams on the right-hand side showing the shape of the localizing solutions illustrate what we have called the *pinch and stretch* phenomenon [van der Heijden *et al* 2002]. Close to $P_C$, the localized path is a very slow modulation of the helical buckling mode, while as the path approaches $P_M$ it becomes essentially a straight un-deflected line which enlarges suddenly to the periodic helix (as we shall see in Fig 8(d)).

### 3.3 Sudden onset of shock-sensitivity

We can use the results for the isotropic rod in a tube [van der Heijden, 2001] to show the onset of *shock-sensitivity* at the Maxwell load [Thompson & van der Heijden, 2014]. In Fig. 8(a) we show the two falling equilibrium paths, while Fig. 8(b) shows the variation of the energy barrier, $E$, with $m$, for both the localized and helical paths. We can see that the energy barrier for escape, highlighted in red, changes suddenly at the Maxwell load, $P_M$, from the high-value associated with the periodic solution ($\alpha_5$, for five wavelengths) to a much lower value governed by the localized curve $\lambda$. Fig. 8(c) shows to the same vertical scale the known diagram for the cylindrical shell.

It is interesting to observe that the $\lambda$ barrier is equal to $E^*$ at the point of collision. The reason for this 'residual' barrier is illustrated in Fig. 8(d). This shows that $E^*$ is in fact the energy of the transition between the zero-energy-density straight solution and the zero-energy-density helical solution. This explains the 'anomaly' mentioned in §3.1.

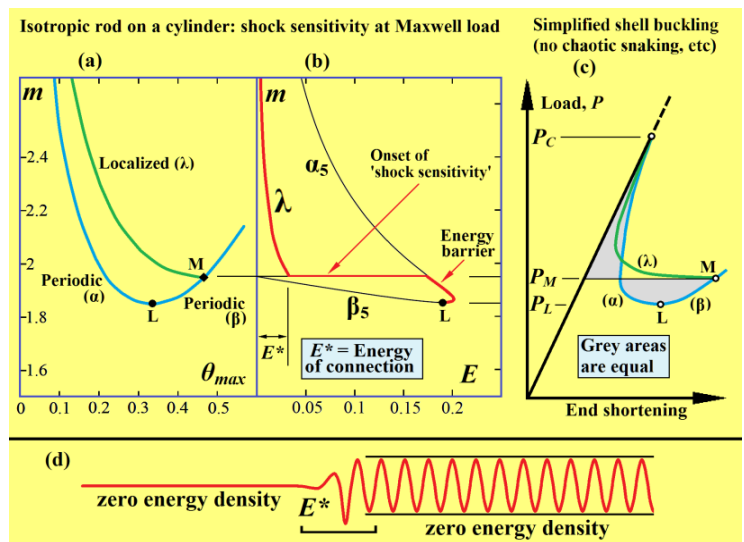

Fig. 8 The onset of shock sensitivity at the Maxwell load for an isotropic rod in a tube, placed alongside the classic shell buckling picture (the latter being drawn here schematically as if it were an integrable system, in contrast to Fig. 16).

## 4. Chaos and Multiplicity in Non-integrable Systems

The elastic strut on a quadratic foundation has been discussed earlier within an integrable perturbation scheme. The complete behaviour is however non-integrable, and the full solution (with no re-stabilization) exhibits spatial chaos. Multiple localized paths offer multiple escape routes: a second example, again without any re-stabilization, is provided by the free, un-constrained, twisted anisotropic rod (namely a rod with differing principal bending stiffnesses).

### 4.1 Spatial chaos in a strut on a nonlinear foundation

A schematic bifurcation diagram showing a representative sample of periodic and homoclinic solutions for the long strut on a quadratic foundation, is shown in Fig. 9, following Buffoni, *et al* [1996].

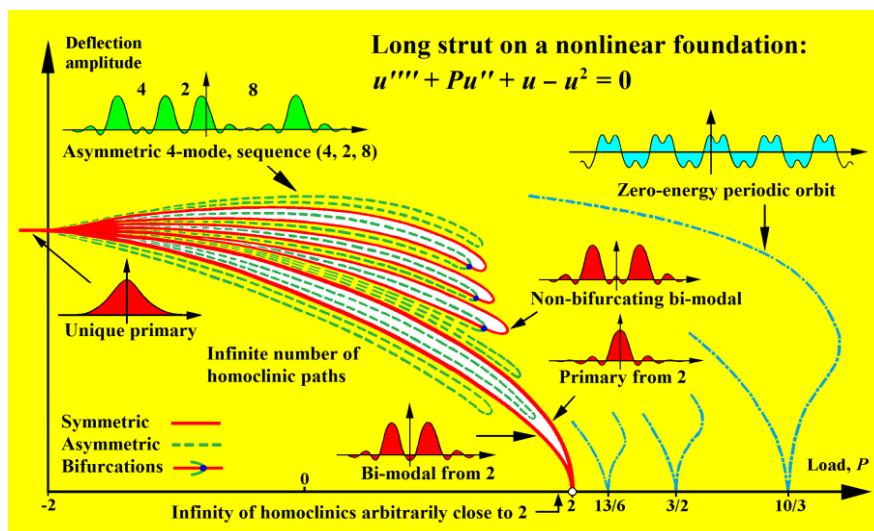

Fig 9. A sketch, based on the results of Buffoni, *et al* [1996], showing details of the multiplicity of localized homoclinic post-buckling solutions exhibited by the strut on a nonlinear quadratic foundation.

The spatial chaos is associated with homoclinic orbits of an equivalent four-dimensional dynamical system. The buckling load is at $P = P_C = 2$, and there are four complex eigenvalues for $-2 < P < 2$. Emerging from the trivial solution at the buckling load are two paths of spatially-symmetric homoclinics, a primary path with a single hump, and a bi-modal with two humps. Over the range of the complex eigenvalues there is an infinite number of homoclinic paths, an infinite number of which approach arbitrarily close to $P_C$. For $P < -2$, we are left with just a single unique primary solution. The symmetric multi-modal orbits exhibit limit points (folds) under increasing $P$. There also exist asymmetric multi-modal paths, and certain of these bifurcate from the symmetric modes immediately before the folds. The diagram also shows a number of significant periodic orbits bifurcating at higher values of $P$.

## 4.2 An infinite number of escape routes

On the right-hand side of Fig. 10 we sketch how these homoclinic solutions can offer an infinity of escape routes from the trivial solution for $P < P_C$. Meanwhile, on the left are two sample solutions derived for the long twisted anisotropic rod [Champneys & Thompson, 1996].

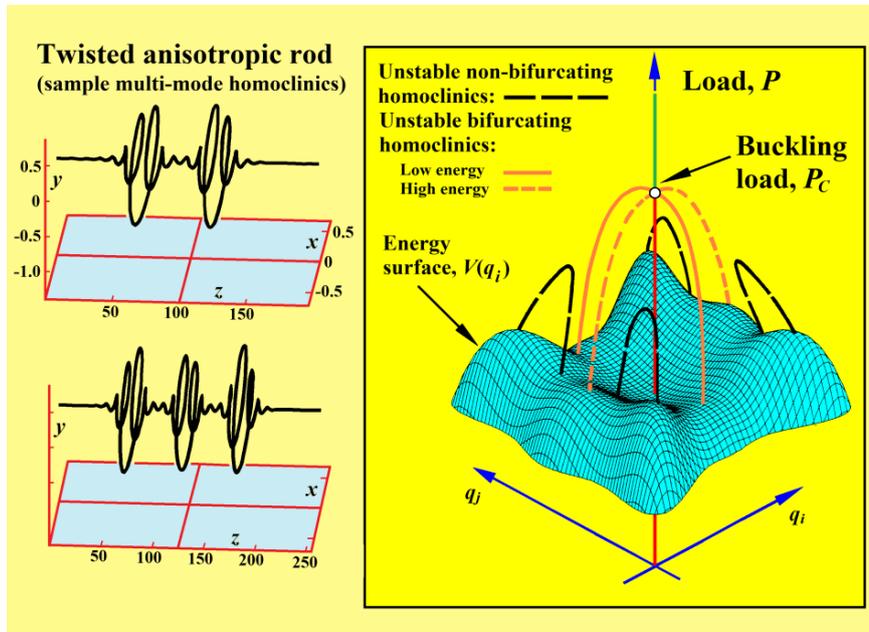

Fig. 10 On the left are shown sample homoclinic solutions calculated for a long pulled and twisted anisotropic rod by Champneys & Thompson [1996]. On the right, we sketch how these localized homoclinic solutions could offer an infinity of escape routes from the trivial solution when subjected to lateral disturbances.

The displayed total potential energy surface is a notional schematic graph of $V(q_i)$ where the $q_i$ are $n$ generalised coordinates describing the deformation of the structure. This surface relates to a fixed value of the load, $P < P_C$.

## 5. Re-stabilization in Non-integrable Systems

We now arrive at our goal of looking at re-stabilization phenomena in non-integrable systems, after which we are in a position to turn our full attention to shell buckling, which has both of these features.

### 5.1 Heteroclinic tangling and snaking paths

We have seen that the end-point of a falling localizing path in an integrable system is governed by a heteroclinic saddle connection in an equivalent oscillator. Now it is well known that when an integrable system is somehow driven into a non-integrable condition (by the addition of extra terms, driving, etc) a heteroclinic connection is smeared out in parameter space into a heteroclinic tangle. This creates chaos and in particular a snaking of the primary localized paths. We look first at the form of this tangling, and then examine how it influences post-buckling curves for a re-stabilizing strut model and for an anisotropic rod in a tube.

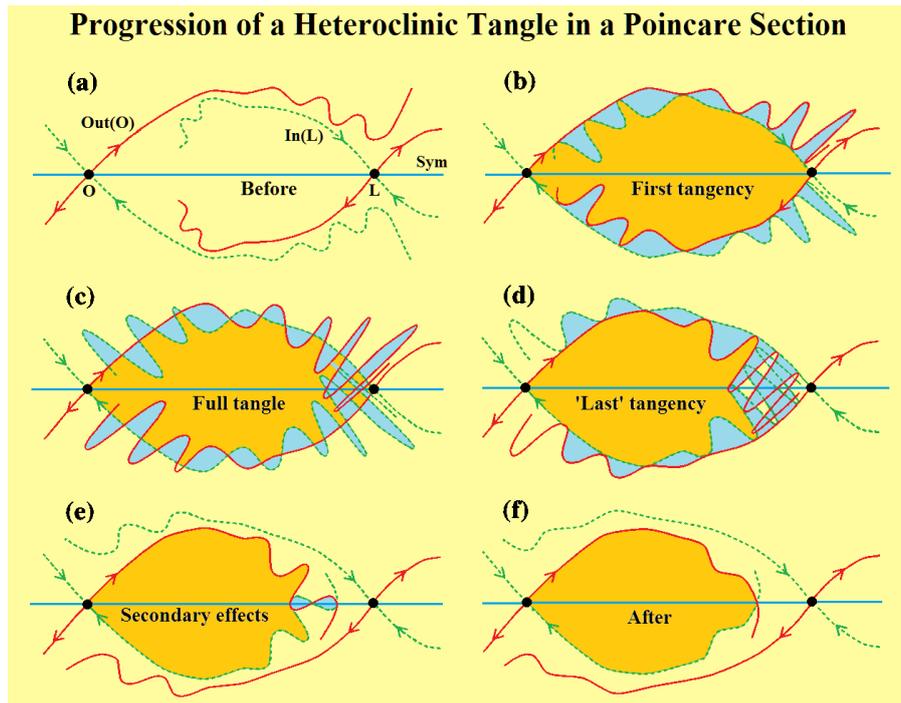

Fig. 11 The smeared-out heteroclinic tangle of a non-integrable system that replaces the heteroclinic saddle connection of an integrable system. Adapted from the paper by Woods & Champneys [1999].

To understand a heteroclinic tangle, we must examine the equivalent dynamics in a Poincaré section, and we do this in Fig. 11, as illustrated in the paper by Woods & Champneys [1999]. Here we show a series of sections as the primary control parameter is slowly varying through the tangle: for the integrable system, all of these pictures would be squeezed to one particular value of the control. We must remember, also, that a manifold drawn in a Poincaré section corresponds to points mapping (effectively jumping) along the curve. This allows curves to cross (as cannot happen in the flow of a phase portrait): and if two curves cross once they must cross an infinite number of times, since forward and backward iterates of the mapping remain on each manifold. The pictures show the first and last tangencies, and then the crossings of the red outsets (unstable manifolds) with the green insets (stable manifolds) of the two fixed points, corresponding to the trivial and periodic states.

## 5.2 Snaking in a strut on a polynomial foundation

The tangling that we have just described gives rise to the snaking of the localized post-buckling path about the Maxwell load, as illustrated in Fig. 12 for a strut on a re-stabilizing polynomial foundation, based on results of Budd, *et al* [2001].

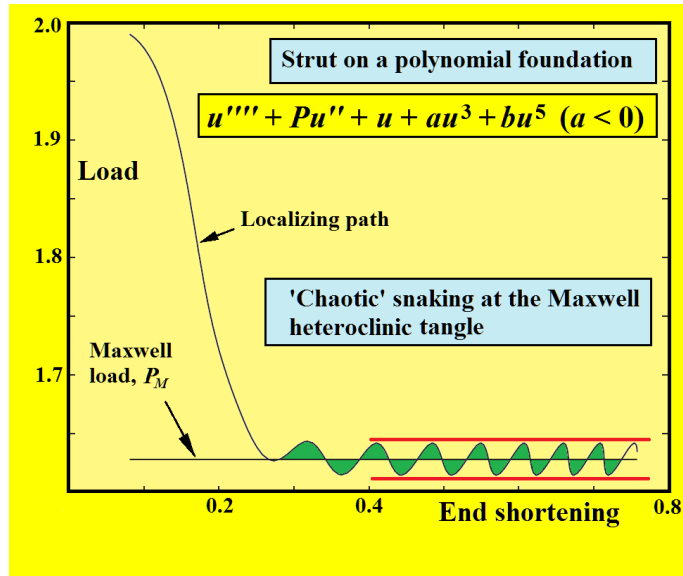

Fig. 12 The sub-critical localized post-buckling curve of a strut on a nonlinear (polynomial) foundation, showing the snaking about the Maxwell line. This snaking can continue to infinity because the strut is assumed to have infinite length. Adapted from [Budd, *et al* 2001].

We see that as the end-shortening increases, the snaking path oscillates about the Maxwell (energy criterion) line. Energy changes on the snaking path are easily monitored by the green areas on this plot of load against the corresponding deflection. The strut is assumed to be very long (effectively infinite), so the end-shortening can and does increase 'indefinitely'. As it increases, we can see how the snaking fits increasing well between a pair of horizontal red lines which correspond to the first and last tangencies of the heteroclinic tangle illustrated in Fig 11.

### 5.3 Snaking for an anisotropic rod in a tube

Our second illustration of heteroclinic snaking is for a twisted anisotropic rod in a cylinder or tube [van der Heijden, *et al*, 2002], as shown in Fig. 13.

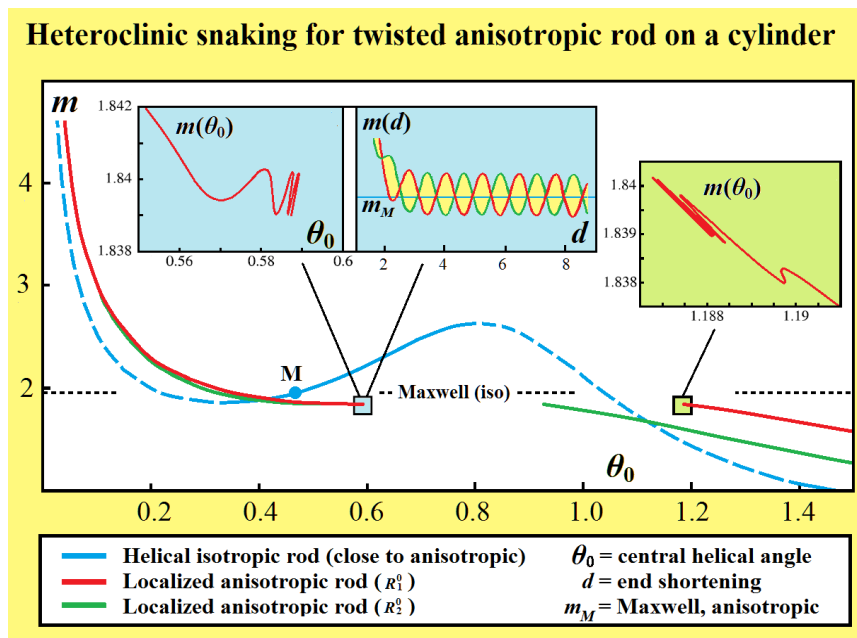

Fig. 13 The end of one (two in the second thumbnail) localized post-buckling path for the non-integrable anisotropic rod in a tube, adapted from [van der Heijden, *et al*, 2002]. Also shown, for comparison, is the equivalent behaviour of the integrable isotropic rod. Notice that unlike the infinitely-long snaking on an end-shortening plot, $m(d)$, the snaking winds to a halt on the central angle plot, $m(\theta_0)$.

The main graph shows a plot of the moment parameter, $m$, against the central Euler angle, $\theta_0$. The helical path shown in blue is drawn for the *isotropic* rod (falling from infinite $m$ as we have described earlier), but the path for the anisotropic rod is very close to this. The Maxwell point for the isotropic rod is denoted by M. Two localized homoclinic paths are drawn for the anisotropic rod, one in red and one in green. Thumbnail enlargements show the details of how the localized anisotropic paths terminate. The first is on a plot of $m$ against $\theta_0$, and we see that the path is essentially winding to a halt. The second thumbnail shows $m$ against the end shortening of the rod, $d$. This, as we have just described for the strut model, can snake to infinity because the rod is assumed to be infinite in length. The two localized paths, green and red, are seen to oscillate (out of phase) about the Maxwell load of the anisotropic system, $m_M$. Meanwhile, the third thumbnail shows the termination of the large amplitude homoclinic, which is beyond the scope of our present discussion.

## 6. Snakes and Ladders

As every child knows, where there are snakes there are usually ladders: and this is the case here in our non-integrable restabilizations.

### 6.1 Asymmetric localization in Swift-Hohenberg equation

To illustrate these ladders we draw on results [Burke & Knobloch 2007, Beck, *et al*, 2009] derived for the Swift-Hohenberg equation, shown in Fig. 14. This is a basic archetypal equation much used by applied mathematicians studying fundamental problems of pattern formation and Turing instabilities: in the form written here it is relevant to the study of nonlinear optics.

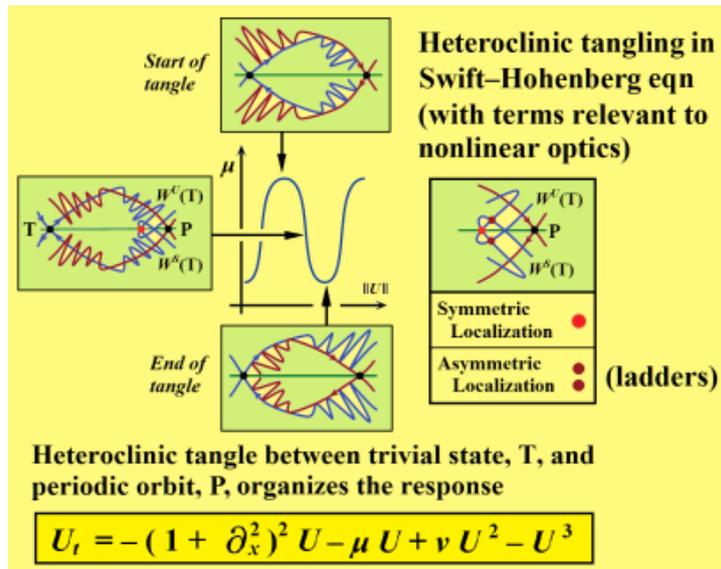

Fig.14 A diagram relating the heteroclinic tangling to the parameter-deflection plot for the Swift-Hohenberg equation in nonlinear optics. Adapted from [Beck, *et al*, 2009].

Once again, we focus on the snaking of a localized homoclinic equilibrium path, now in the space of the control parameter, $\mu$ and a measure of the deflection $\|U\|$. This is related to the three (complete) Poincaré sections in which the progression of the heteroclinic tangling is illustrated. The enlargement on the right, shows details of the fully developed tangle. In the latter we can identify a symmetric localization shown is red, and two nearby asymmetric localizations shown in brown.

## 6.2 Ladder rungs connecting symmetric snakes

The grouping of localizations that we have just seen hints at the presence of the asymmetric ladders that we show next in Fig. 15.

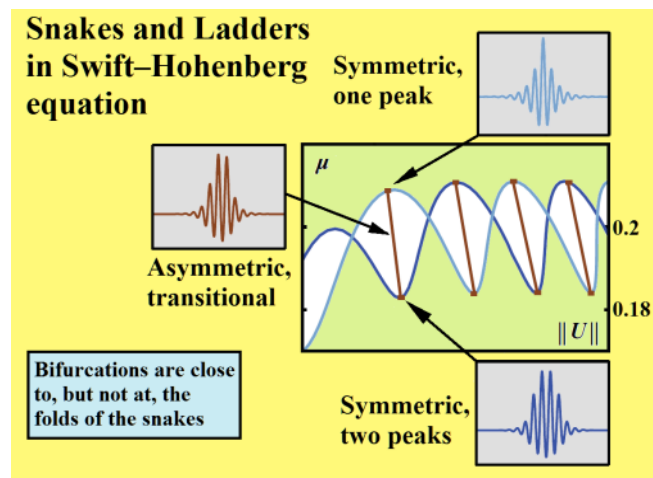

Fig. 15 An illustration of the ladder rungs that connect two snakes in the Swift-Hohenberg equation. These are paths of asymmetric solutions that bifurcate from the symmetric solutions, allowing a transition between the one-peak and two-peak curves. Adapted from [Beck, *et al*, 2009]. Similar rungs might be expected in other snaking solutions, in, for example, the strut on a polynomial foundation.

Here, for the same equation, we show the calculated paths [Beck, *et al*, 2009] of two symmetric homoclinics, one with a single peak, and the other with two peaks. Finally the paths of asymmetric homoclinics form the transitional rungs of a ladder, bifurcating from the symmetric solutions. Note that the bifurcations are close to, but not at, the folds of the snakes, as we could deduce from the enlarged Poincaré section in the previous figure.

## 7. Theoretical phenomena in compressed cylindrical shells

When discussing or analysing localization phenomena in the long axially compressed cylindrical shell, the localization is usually taken to be in the axial direction only, as in Fig. 16 adapted from [Hunt, 2011]. Meanwhile there is assumed to be a fixed number of circumferential waves ($n$ =11 in the figure).

Note also that most of the work described here is based on the *von Karman–Donnell* shell equations. These are only valid for relatively small nonlinear deflections, and probably will not give accurate results for the snaking phenomena. After emphasizing this caveat, it must be remembered that what I say in this theoretical section about the 'behaviour of a cylindrical shell' is more accurately described by the phrase 'the behaviour of the von Karman-Donnell equations'.

### 7.1 Snaking and progressive localization

A long cylindrical shell under axial compression is a non-integrable problem and as such exhibits the afore-mentioned snaking phenomena. Results by Giles Hunt and his co-workers [Lord *et al* 1997; Hunt *et al* 1999; Hunt 2011] are summarised in Fig. 16. This displays the snaking of a localized homoclinic solution, due to the chaotic heteroclinic tangling that occurs at the Maxwell load, $P_M$.

The circumferential wave number (the number of full waves in the circumference) is fixed at $n = 11$, while the radius to thickness ratio of 405 corresponds to one of Yamaki's experiments [Yamaki 1984]. The localized curve which has fallen sharply from the linear bifurcation value of $P_C$ is seen to snake about the Maxwell load, and the deformation patterns corresponding to points A, B and C on the curve are illustrated on the left-hand side of the figure (these three points were chosen only for illustration purposes, and have no other significance). These patterns show how the extent of the axial localization increases as we travel along the snaking path.

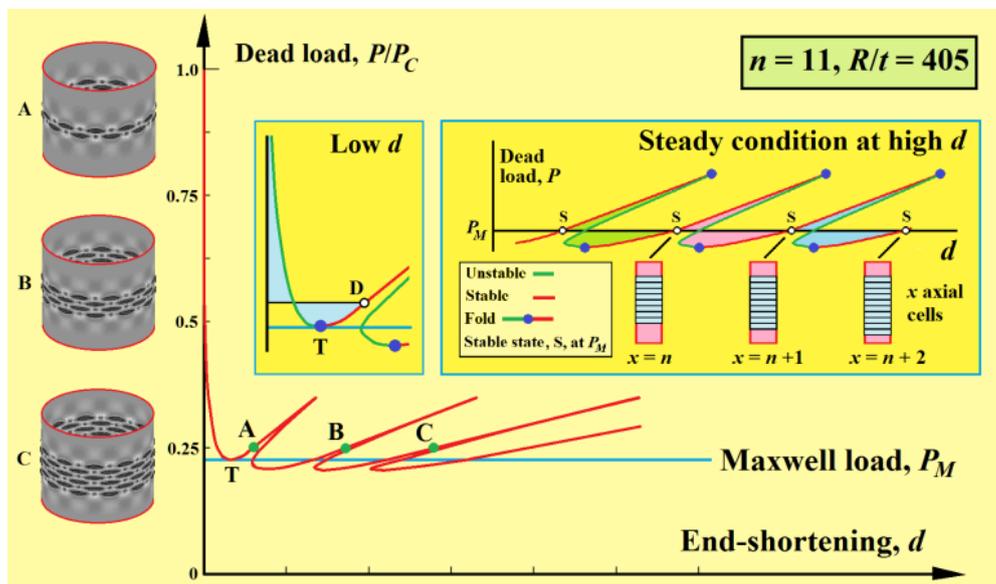

Fig. 16 A localized sub-critical post-buckling solution for the axially compressed cylindrical shell, exhibiting snaking about the Maxwell load. Notice that the trivial contribution has been subtracted from the plotted end-shortening. Pictures of the three computed cylindrical deformations correspond to points A, B and C on the equilibrium path. Adapted from the references given in the main text.

The first insert shows a magnified region of the post-buckling curve for low $d$. Here, we see an interesting event at point D, where the two blue areas are equal: this corresponds to what we might call a secondary Maxwell point relating to the localized path. The apparent tangency between the post-buckling curve and the Maxwell line, at point T, is a coincidence for this particular shell, with no general significance.

The second insert shows a magnification of the curve for high $d$. Under the dead load, $P$, the stability of the snaking path changes at the marked folds, but there may well be (so-far undetected) bifurcations, like those in the Swift-Hohenberg equation. As $d$ increases towards infinity, the behaviour settles into a fixed pattern (though the tilt of the waves continues to increase), and this is reinforced by the physical behaviour in which the localized cells spread outwards as reflected in the shapes at A, B and C. In this final steady progression, the total potential energy (strain energy minus $Pd$) of the stable states, S, can be expected to tend towards that of the trivial state at $P_M$. If this were the case, adjacent areas such as those in green, would become equal in the infinite limit.

Notice that when we are in the smeared-out heteroclinic regime of $P$, the (static-dynamic) analysis picks out from the tangle at every $P$ a homoclinic orbit with special features (such as one-hump). In this way we extract a continuous snaking path corresponding to these special features from the chaos of the tangle. Prescribing different 'special features' we would get different paths (akin to all the paths of Fig. 9 uncovered by Buffoni, *et al* [1996]).

### 7.2 Lowest mountain-pass against buckling

We have seen in a number of examples that a localized equilibrium path falling from the linear critical buckling load, $P_C$, yields a *lower* energy pass than the falling periodic solutions. In a very significant direct attack on the problem of the minimum energy barrier Horak *et al* [2006] use sophisticated algorithms of mathematical analysis (including the mountain pass theorem of Ambrosetti & Rabinowitz [1973]) to determine the *lowest* energy barrier against disturbance-induced buckling for a long cylindrical shell under axial compression. This corresponds to a precise localized state, and their results are shown in Fig. 17. Here the main graph shows how the energy barrier of this state varies with the axial load on the cylinder. The same path of states is shown in the green thumbnail on a plot of $P/P_C$ against a measure of the deflection. Meanwhile two views of the determined shape of the unstable localized state are shown on the right hand side of the figure, this shape depending on the thickness to radius ratio of the shell. This is a very significant result, which will form the basis of our later suggestions for new experimental work on shell buckling.

We note, finally, that in this study the shell is constrained to an *average* end-shortening, meaning that the ends of the cylinder are free to tilt, thereby accommodating a circumferentially localized dimple. This is quite different from conventional shell testing, where the rigid platens supporting the ends of the shell would normally be constrained to remain perpendicular to the axis of the cylinder.

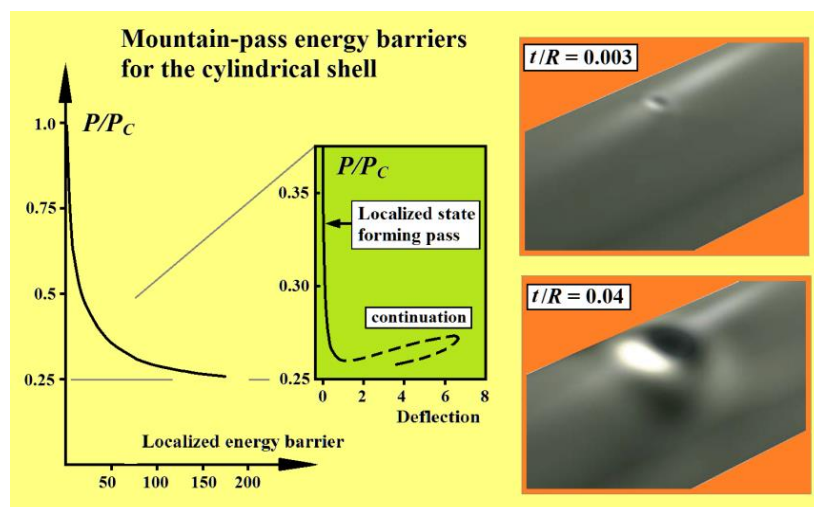

Fig. 17 Results of Horak *et al* [2006] who determined the localized solution that gives the lowest mountain-pass escape barrier for a compressed cylindrical shell. Notice that unlike the pictures of Fig.16, the displayed displacement is localized not only axially, but also circumferentially. It is, moreover, symmetric in both of these directions

## 8. Cylinder Experiments Old and New

### 8.1 Historical scatter of experimental results

As a background to the following ideas about new experiments on shells, it is useful to take a quick look at the historical collapse loads, many of which were obtained in the first half of the twentieth century. The points shown in Fig. 18 are taken from the paper by Seide *et al* [1960]. They show the variation of the experimental collapse loads, $P_{EXP}$, as a fraction of the classical linear theoretical values, $P_C$, against the radius to thickness ratio of the test specimens, $R/t$. This graph also shows a curve based on the old empirical knock-down factor recommended by NASA for design purposes.

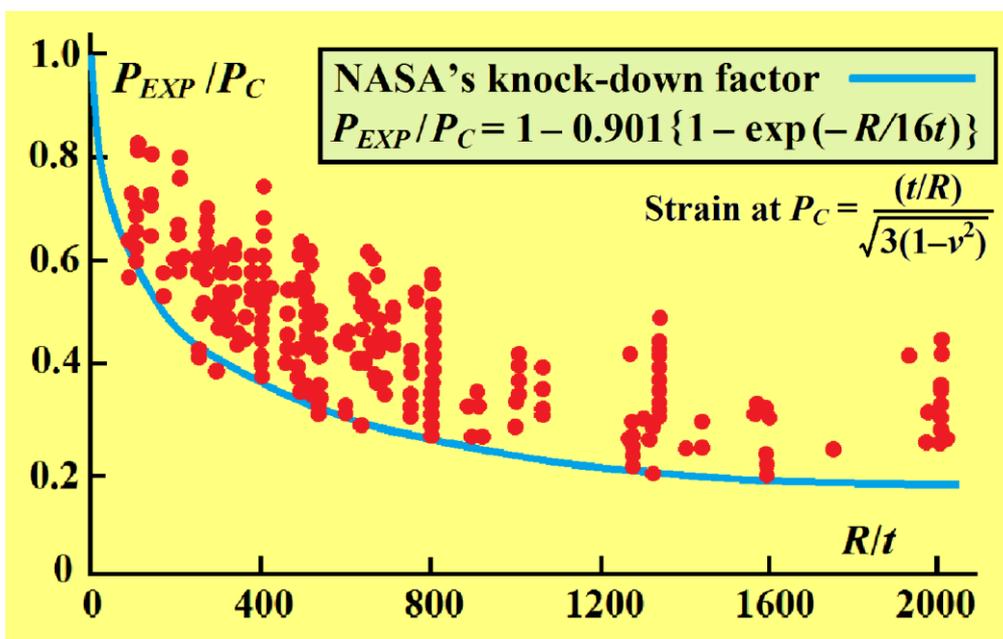

Fig. 18 Historical scatter of experimental buckling loads for an axially compressed cylindrical shell, adapted from Seide *et al* [1960]. The blue curve represents the NASA knock-down factor displayed in the box.

Similar (often the same) results taken from Brush & Almroth [1975] are displayed in a log-log plot introduced by Chris Calladine and his co-workers [Calladine & Barber 1970, Mandal & Calladine 2000, Zhu et al 2002] in Fig. 19(a).This shows, very convincingly, that the best fit for these experimental results has the log-log slope of −1.48. In a series of papers, they give a number of convincing reasons for this slope of approximately −1.5, based on careful experiments on the self-weight buckling of a standing cylinder, with a free unloaded top; together with some supporting theoretical results.

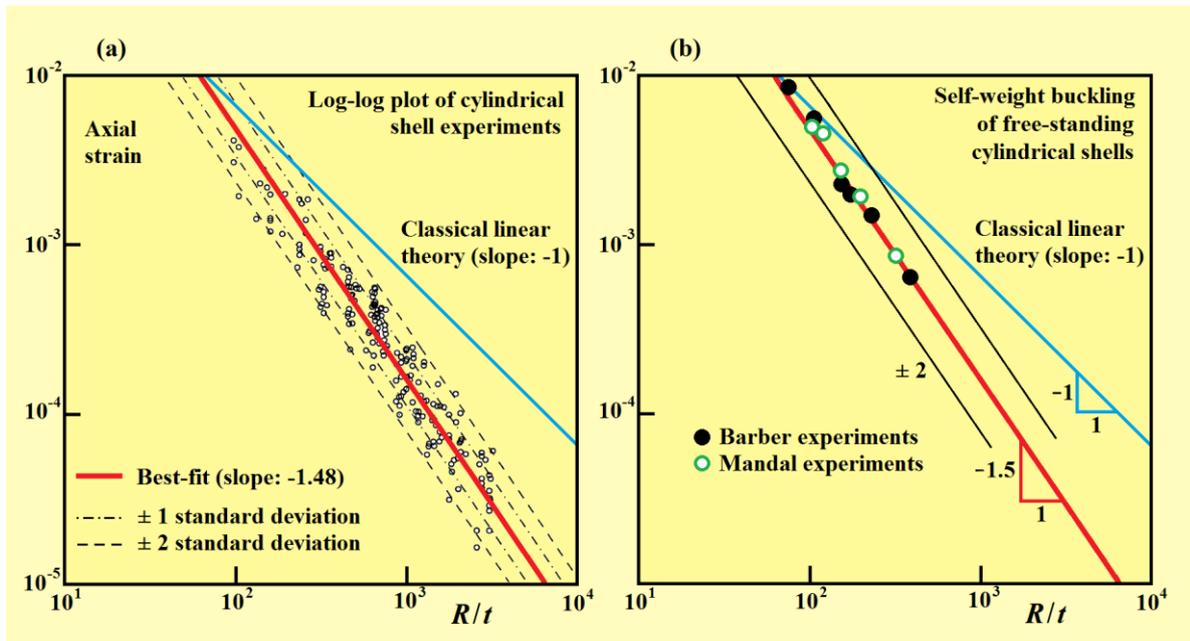

Fig. 19 (a) shows the experimental buckling loads for an axially compressed cylindrical shell, on a log-log plot. On (b) is an identical log-log plot showing self-weight buckling tests conducted by Barber and Mandel. Adapted from the studies of Chris Calladine and his associates [Mandal & Calladine 2000, Zhu *et al* 2002].

Results of Calladine's extended experimental studies of the self-weight buckling of free-standing open-topped cylindrical shells are shown in Fig. 19(b). They show remarkable consistency, and lie almost exactly on the best-fit line of Fig. 19(a) with slope approximately −1.5. Together with some theoretical arguments, Calladine uses these and other experiments to argue that the wide scatter of the axially compressed shells is due to their 'statical indeterminacy' (which contrasts with the free shells under self-weight loading). His argument uses the analytical "concept" of an inextensional dimple with an elastically strained boundary, which provides a satisfactory explanation of the self-weight data. Moreover, he recorded [Lancaster *et al*, 2000] a particularly high experimental buckling load with a cylinder ($R/t = 2000$) by introducing boundary conditions (end discs fastened with frictional clamps) that significantly reduced self-stress.

## 8.2. NASA's current research programme

It is interesting to observe that, some 70 years after the key papers of von Karman & Tsien [1939, 1941], shell buckling and post-buckling are still of major concern to NASA in their design of deep-space rockets. Indeed, shell buckling is the primary factor in the design of thin-walled launch-vehicle structures that must carry compressive loading. The core stage design of the current *Space Launch System* (SLS) is, for example, completely driven by buckling, so unduly conservative design factors which increase structural mass must be avoided. This is emphasized by the fact that the space shuttle $LH_2$ tank was tested to a load greater than 140% of its design load.

It is hardly surprising, then, that NASA is currently running a programme of full scale tests on stiffened shells under mixed loading as illustrated in Fig. 20. This programme seeks a rational way to replace the existing reliance on knockdown factors based on historical experimental data. The pedigree of this data (often from the period 1920 – 1960) is difficult to assess, and many of the tests are not relevant to modern launch vehicle constructions. In particular, most NASA shells are stiffened, and so are less sensitive to imperfections than their unstiffened counterparts.

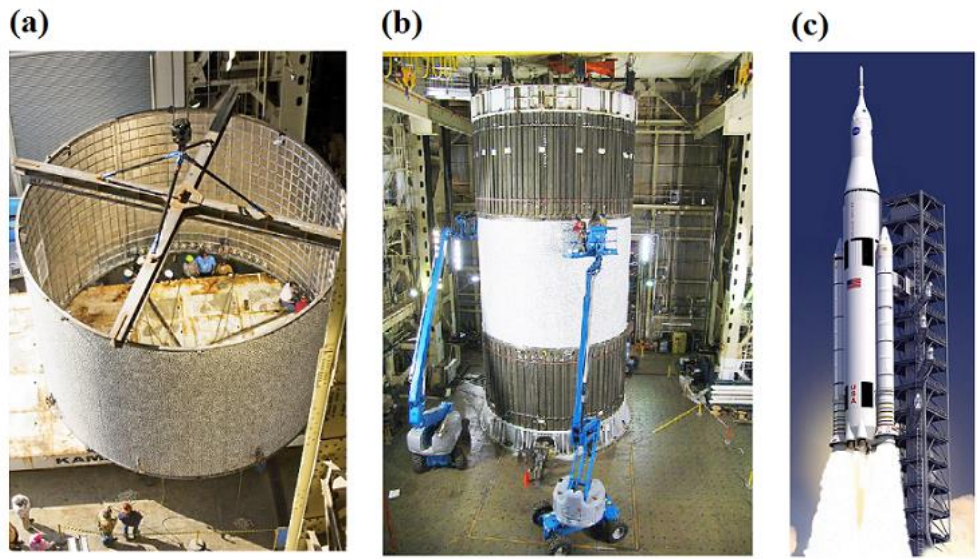

Fig. 20 (a) and (b) illustrate NASA's full scale test in 2013 of an unused aluminium-lithium shell left over from the Space Shuttle programme. At 8.3m diameter, it parallels the fuel tank (namely the structural skin) of the SLS. The stiffened and pressurized shell was compressed to an explosive buckling failure in a rigid loading device. Dots on the cylinder allowed 20 high-speed cameras to record minute deflections under a load of circa $10^6$ pounds force. Shown in (c) is an artist's impression of NASA's Space Launch System, a new heavy-lift vehicle standing 61m tall with an 8.5m diameter. Reproductions courtesy of NASA.

Experiments are being made on full-scale shells, left over from the Space Shuttle, but we should note that these are relatively short (compared to the diameter), they are stiffened, and the applied loading is complex, rather than just uniform compression. One fact that NASA hopes to establish from this programme is that modern computer codes can accurately predict the correct experimental failure loads of their full-scale experimental shells once they are given the known measured imperfections. With advances in computations and testing techniques, the principal investigator, Mark Hilburger, says that weight reductions of about 20% are now confidently expected in the design of the projected Space Launch System (SLS) which is due to fly in 2017. More information about this work can be found in [Hilburger, 2013].

## 9. New Experimental Approach to Shock-Sensitivity

The shape of the lowest energy barrier determined by Horak *et al* [2006] for an axially loaded cylindrical shell (Fig. 17) looks remarkably like the small dimple that might be pressed into the cylinder by a researcher's finger. This immediately suggests a new form of experimental test on a compressed shell (of any shape) in which a lateral point load is applied by a rigid loading device. This would seem to be a useful type of non-destructive and non-invasive test for a shell to determine its shock-sensitivity.

## 9.1 Rigid lateral probe

The type of test configuration that we have in mind is illustrated, for a cylindrical shell, in Fig. 21 where the lateral 'probe' moves slowly forward along a fixed line driven by a screw mechanism.

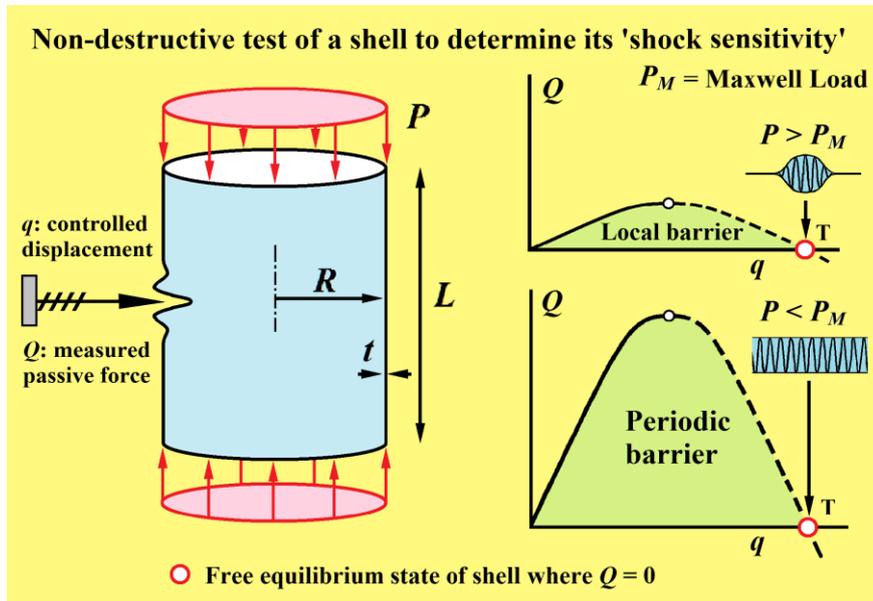

Fig. 21 An impression of the proposed experimental procedure in which a rigid probe is used to construct a lateral-load versus lateral-displacement graph, $Q(q)$. This graph ends with $Q = 0$ at a free equilibrium state of the shell, and the area under the curve gives the corresponding energy barrier. Note that at $Q = 0$ the rigid probe is stabilizing a state that would otherwise be unstable for the free shell.

Here we have the controlled displacement, $q$, producing a passive reactive force from the shell, $Q$, which is sensed by the device, giving finally the plot of $Q(q)$. This is all to be done at a prescribed value of the axial compressive load, $P$, which might itself be applied in either a dead or rigid manner. In the simplest scenario, the $Q(q)$ graphs might be expected to look like those sketched on the right-hand side, the top for $P$ greater than the Maxwell load and the lower one for $P < P_M$.

When the test reaches the point T, where $Q = 0$, we have located a free equilibrium state of the shell, hopefully, the desired lowest mountain pass. It is interesting to note, here, that Takei *et al* [2014] use an imposed lateral displacement in their computations to find the unstable Maxwell state of a thin film. The shapes drawn as thumb-nails in the figure are purely notional, and the shape that we would hope to find is that of Horak, shown in Fig. 17. The latter shape depends on the thickness to radius ratio, $t/R$, and it gives us some intuitive feeling for whether the deformation will be within the elastic range of a given shell. If we finally evaluate the area under the $Q(q)$ curve, this will give us the energy barrier that has to be overcome to cause the shell to collapse at the prescribed value of $P$.

Notice that if the curves have the forms drawn (with no folds or bifurcations), we can be sure that the shell will remain stable up to T under the controlled $q$. If the probe is glued (welded or fastened) to the shell, we will be able to pass point T, with the probe then carrying a negative $Q$: if however the probe is just resting against the shell, a dynamic jump from T will be observed, probably damaging the shell due to large bending strains. Clearly gluing is preferred, to prevent this jump, but other factors must be considered. On the negative side, fixing the probe to the shell may itself cause damage, and may restrict the free deformation that we are seeking to find. In particular, it might also restrict the shell by preventing a rotational instability. Perhaps the ideal solution would be to have an 'equal and opposite' second probe inside the shell at the same point, moving at the same rate as the probe outside.

## 9.2 Bifurcations and the need for control

Now under the lateral point load, the *initial* deflection will be symmetric in both the axial and circumferential directions, and we should note that all of Horak's lowest mountain passes (localized saddle solutions) have both these symmetries. However, the story might not be as simple as we have so far suggested. Three features that could be encountered in the $Q(q)$ curves are as follows. The first possibility is a vertical fold, at which $dQ/dq =$ infinity, from which the combined system could jump at constant $q$, with unknown outcome. The second and third are symmetry-breaking pitchfork bifurcations that could break one (or subsequently both) of the initial symmetries. A bifurcation that is sub-critical (involving asymmetric equilibrium states at a lower value of the controlled displacement, $q$) would give a jump to an unknown state: this is akin to the elastic arch under dead loading that we shall examine in §9.3. Meanwhile a super-critical event would give a new path which could possibly be followed in the test. Note that these symmetry-breaking bifurcations would generate a new mode of deformation in the shell.

To overcome any such instabilities, it would be necessary (as we shall demonstrate in Fig. 25) to supplement the main probe by one or more probes that are systematically adjusted until all their reactive forces are simultaneously zero. Jan Sieber has, for example, developed non-invasive control methods which can follow an experimental system as it is loaded into what would otherwise be an unstable regime. He has applied these methods successfully to a number of systems at Bristol University [Barton & Sieber, 2013].

## 9.3 Analogy with a deep elastic arch

The symmetry-breaking that we have discussed is indeed well-known in the response of a deep (as opposed to a shallow) elastic arch under *dead* loading [Thompson 1975, Thompson & Hunt, 1973, 1983]. This bifurcation allows the arch to woggle-through to a large-deflection symmetric state avoiding the high energy penalty of remaining symmetric throughout, as is illustrated in Fig. 22. An exactly similar figure arises in the tensile instability of the atomic lattice of a close-packed crystal, governed by Lennard-Jones potentials, where the bifurcation triggers a symmetry-breaking shearing mode of deformation [Thompson 1975; Thompson & Shorrock 1975].

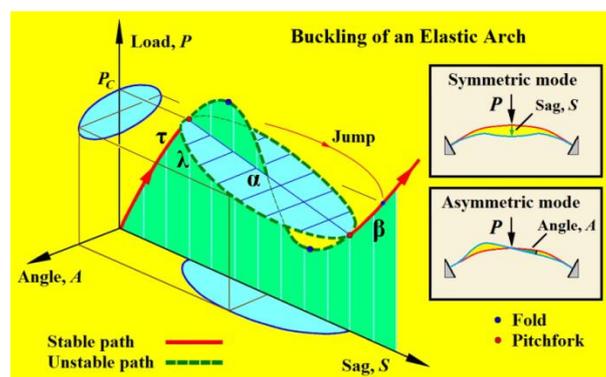

Fig. 22 The buckling and snap-through of a relatively deep arch under a dead load, *P*, shows how a bifurcation into an asymmetric mode allows the arch to woggle through, avoiding the large energy penalty of remaining symmetric. Notice the schematic curved dynamic jump at the constant load, $P_C$, that carries the arch to its re-stabilized symmetric state. The associated energy contours are show in Fig 23.

## 9.4 Localized saddle allows escape past a mountain

This response of a deep arch answers another question that comes to mind when thinking about the mountain pass of Horak, *et al* [2006]. How can it be that escape over a spatially-localized energy pass can allow the shell to collapse to very large amplitudes, seemingly getting past the high periodic barrier? Well this is very similar indeed to what happens in the deep arch. If we sketch the contours of the total potential energy of the arch, *V*(*A*, *S*) at a value of the load, *P*, just less than the bifurcation load, $P_C$, we get a diagram similar to that of Fig. 23.

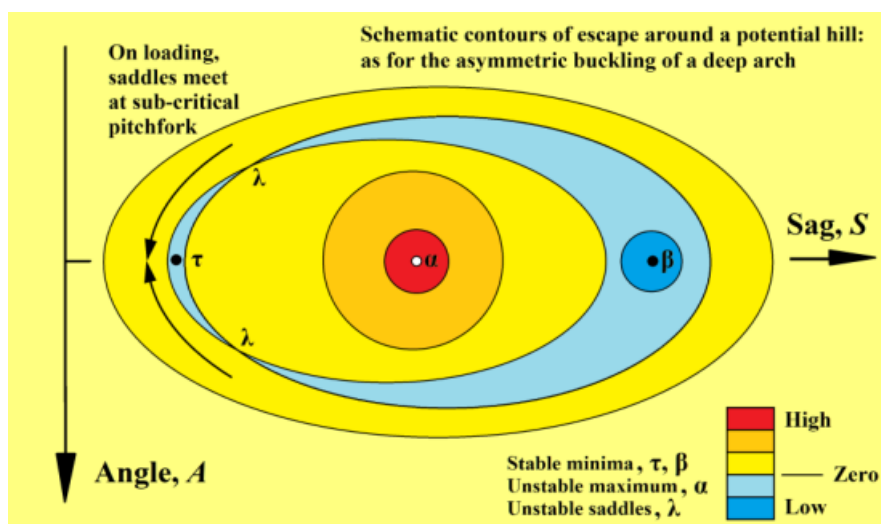

Fig. 23 A sketch of the total potential energy contours of the deep arch at a value of the load just less than $P_C$ (shown in Fig. 22). At the sub-critical bifurcation, when the saddles, λ, collide with the symmetric un-buckled state, τ, the arch can jump dynamically along the falling blue valley to β, avoiding the high energy peak, α.

Here we see that as *P* approaches $P_C$ the two asymmetric saddles, λ, move towards and collide with the symmetric state τ. The arch can then escape along one of the falling blue 'channels', the one chosen depending on small disturbances and imperfections. In this way the system gets around the high potential energy barrier represented by the symmetric hilltop, α.

## 9.5 Successful simulation with a single probe

To examine, in a theoretical context, the use of our proposed experimental procedure my colleague Jan Sieber has just completed a feasibility study summarised in Fig. 24. In this work he has constructed a computer model of a shell-like structure made of connected links and springs shown on the left-hand side in its deformed state under a dead axial load *P*. We can see that that there is considerable overall Euler buckling of the structure, but this does not detract from its value in establishing the methodology.

It should be emphasized that the shell-like model structure is only a very rough approximation to a continuous shell: just close enough to check out the feasibility of the proposed technique, but no more. It is a dynamic model of a system of particles (with mass and inertia) and springs forming a rectangular mesh which has been bent around to form a cylinder. There are 24 particles on each horizontal ring and 35 rings. So far, in this very preliminary study, there has been no attempt to fit the rotational and extensional spring constants to match those of a continuum shell. However, the model does show all the features of a compressed cylindrical shell in a convincing way.

In the simulations the structure is loaded by a rigid displacement $q$, and the passive force $Q$ is recorded. The first result, from a test at axial load $P = 15.2$ (in arbitrary units), is shown in the top graph of $Q(q)$. This has the expected parabola-like shape, and $Q$ drops continuously to zero at the end of the test, where the shell is essentially free at $q = q_F$. During the simulation, for each recorded equilibrium state, the (real) eigenvalues of the Jacobian are evaluated, and if any one becomes positive an instability has been encountered. To monitor this possibility the maximal eigenvalue is plotted (in red on the top graph) in what will usually be a curve with crossover points, whenever a new mode becomes the 'softest'.

We see that the maximal eigenvalue remains less than zero throughout the first test, but it can be seen to increase rapidly towards the end. This, then, is an example in which the $Q(q)$ test can be performed in a stable manner up to $q_F$, allowing the area under the curve to be evaluated as the required energy barrier.

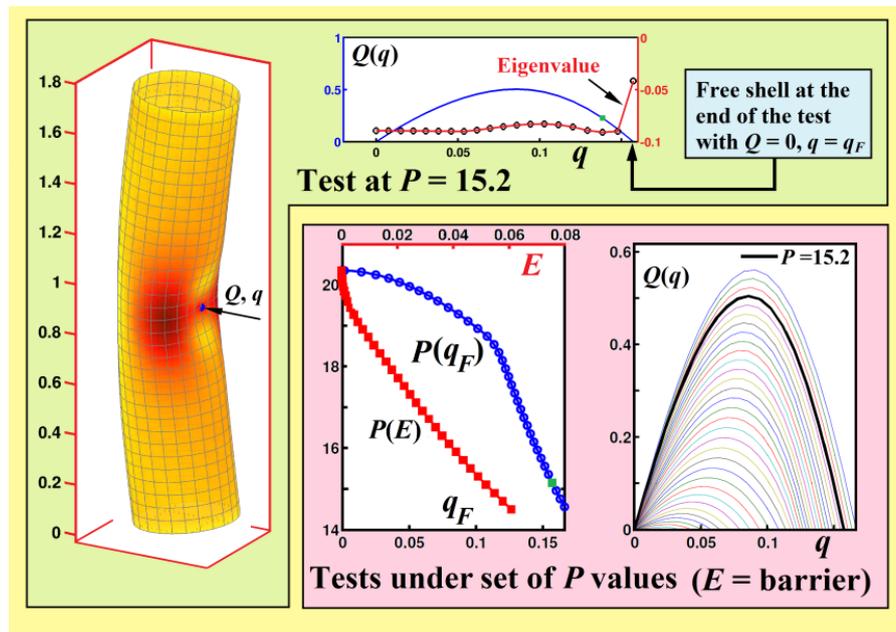

Fig.24 Results of a computer simulation on a shell-like structure showing a case in which a single rigid probe can generate the full $Q(q)$ response, leading to the required relationship between the dead axial compressive load, $P$, and the energy barrier, $E$. The colouring of the shell shows contours of the strain energy density. Results supplied by Jan Sieber (Exeter University) in a private communication.

Tests under a set of $P$ values are summarised in the red box. On the right-hand side we show the $Q(q)$ curves for the full range of $P$ values, with the black curve identifying the one at $P = 15.2$. On the left we show the results as $P$ against $q_F$ and finally as $P$ against the energy barrier denoted by $E$. The sharp and rapid fall of the energy barrier as we reduce the load $P$ from its critical value $P_C$ is our expected result.

## 9.6 Shell simulation with a controlled bifurcation

In simulated tests on this shell-like structure with different stiffness characteristics, Jan Sieber has produced the results shown in Fig. 25.

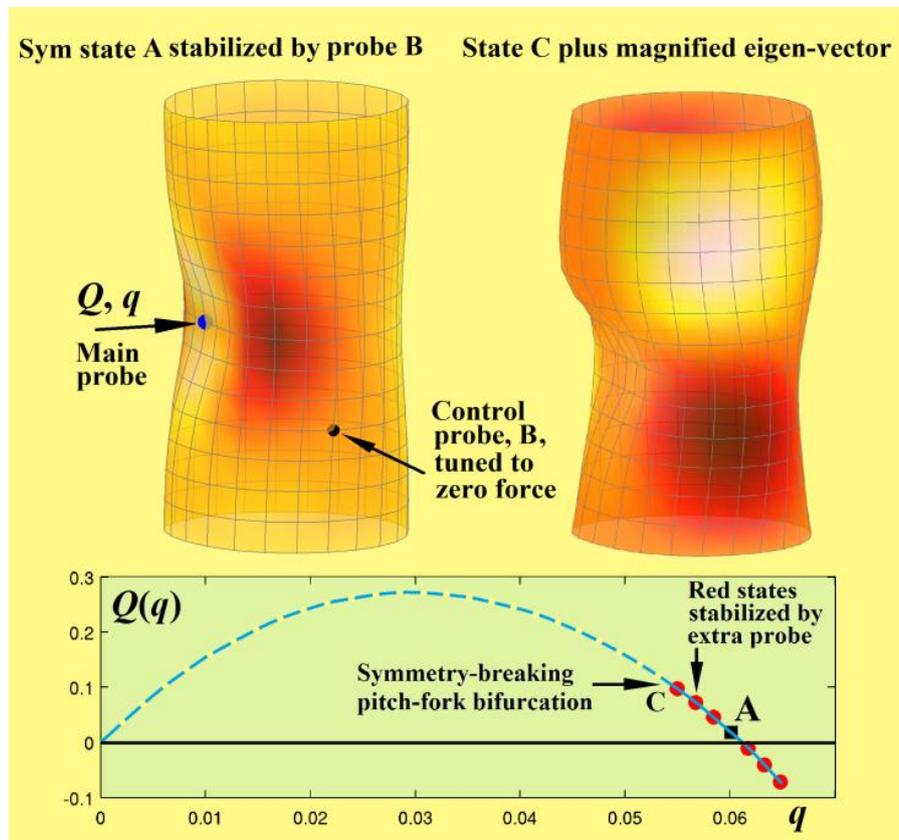

Fig. 25 Control of a symmetry-breaking pitch-fork bifurcation. An example of a simulation in which, under a single probe, the $Q(q)$ path would become unstable at a symmetry-breaking pitchfork bifurcation. This requires the introduction of a second controlled probe, B, placed where the eigen-deflection of the pitchfork is large. With the second probe, the test has been successfully completed. For the shell on the left, the colouring shows contours of the strain energy density; on the right, it represents contours of (one component of) the eigen-displacement. Results supplied by Jan Sieber (Exeter University) in a private communication.

Here the eigenvalues showed that the $Q(q)$ diagram had reached a symmetry-breaking pitchfork bifurcation (whether super- or sub-critical is not known) which required the introduction of a second control probe. This was placed where the critical eigen-vector of the pitchfork bifurcation has a particularly large deflection. The red dots on the $Q(q)$ diagram indicate those states that could only be observed with the use of the extra probe, B. This probe is tuned to provide no force, as we might, for example, demonstrate the second buckling mode of a pin-ended strut by lightly holding the centre between two fingers. One such controlled state is illustrated by the left-hand cylinder. Meanwhile, on the right, is a view of the symmetric state at the pitchfork with the addition of the suitably magnified eigen-vector. This shows that the pitchfork has violated the up-down symmetry (but not the circumferential symmetry). Notice that the deflection curve has been continued past the $Q = 0$, $q = q_F$ free state condition. Albeit conducted on a computer, this is an excellent and encouraging 'proof of concept' for our proposed experiments.

## 10. Concluding remarks

The characteristics of post-bifurcation equilibria that we have described in the theoretical sections of this paper have wide applications in localization studies in solid and fluid mechanics, and in pattern formation by Turing instabilities in chemical and biological kinetics [Dawes, 2010]. Meanwhile, the Maxwell load is appearing in many guises, in for example the recent study by Cao and Hutchinson [2012] of the surface wrinkling of a compressed half-space, which is extremely imperfection-sensitive, with close parallels to the behaviour of the cylindrical shell.

The theory has revealed a sudden onset of 'shock-sensitivity' in the buckling of compressed shells and shell-like structures, which is an important design consideration for real structures in noisy *operational* environments. To detect this experimentally in a model or prototype structure, we have proposed a novel technique of controlled non-destructive testing, which looks particularly promising. No such experiment has been conducted so far, but Jan Sieber's simulations are a very promising 'proof of concept'. His work also demonstrates the need for extra control points, a single one of which successfully inhibited a pitch-fork instability. Looking to the future, it may be that the theoretical techniques employed by Horak *et al* [2006] can be adapted to work in an experimental environment under multiple controls.

### Acknowledgements

I would like to thank the following for comments, suggestions and help during the preparation of this paper: Alan Champneys (Bristol), Allan McRobie (Cambridge), Chris Calladine (Cambridge), David Barton (Bristol), David Wagg (Sheffield), Gabriel Lord (Herriot Watt), Gert van der Heijden (UCL), Giles Hunt (Bath), Jan Sieber (Exeter), John Hutchinson (Harvard), Jonathan Dawes (Bath), Lawrence Virgin (Duke), Mark Hilburger (NASA), Mark Peletier (Eindhoven). A particular impetus came from a small and informal brain-storming session that Giles Hunt organized at Bath this year.

xx Finished manuscript, 10/09/2014 FC